\documentclass[12pt,chicago]{emulateapj}

\usepackage{enumerate}
\usepackage{natbib}
\begin{document}

\title{Probing the Earliest Stage of Protostellar Evolution\\
          --- Barnard 1-bN and Barnard 1-bS}

\author{Yun-Hsin Huang\altaffilmark{1,2,3}}

\author{Naomi Hirano\altaffilmark{3}}

\affil{$^{1}$Institute of Astrophysics, National Taiwan University, Taipei 10617, Taiwan} 
\affil{$^{2}$E-mail: yunhsin@asiaa.sinica.edu.tw}
\affil{$^{3}$Academia Sinica, Institute of Astronomy \& Astrophysics, P.O. Box 23Ð141, Taipei, 106, Taiwan, R.O.C.}

\begin{abstract}
Two submm/mm sources in the Barnard 1b (B1-b) core, B1-bN and B1-bS, have been observed with the Submillimeter Array (SMA) and the Submillimeter Telescope (SMT).
The 1.1 mm continuum map obtained with the SMA reveals that the two sources contain spatially compact components, suggesting that they harbor protostars.
The N$_2$D$^+$ and N$_2$H$^+$ $J$=3--2 maps were obtained by combining the SMA and SMT data.
The N$_2$D$^+$ map clearly shows two peaks at the continuum positions. The N$_2$H$^+$ map also peaks at the continuum positions, but is more dominated by the spatially extended component.
The N$_2$D$^+$/N$_2$H$^+$ ratio was estimated to be $\sim$ 0.2 at the positions of both B1-bN and B1-bS.
The derived N$_2$D$^+$/N$_2$H$^+$ ratio is comparable to those of the prestellar cores in the late evolutionary stage and the class 0 protostars in the early evolutionary stage.
Although B1-bN is bright in N$_2$H$^+$ and N$_2$D$^+$, this source was barely seen in H$^{13}$CO$^+$. This implies that the depletion of carbon-bearing molecules is significant in B1-bN.
The chemical property suggests that B1-bN is in the earlier evolutionary stage as compared to B1-bS with the H$^{13}$CO$^+$ counterpart. The N$_2$H$^+$ and N$_2$D$^+$ lines show that the radial velocities of the two sources are different by $\sim$ 0.9 km s$^{-1}$.
However, the velocity pattern along the line through B1-bN and B1-bS suggests that these two sources were not formed out of a single rotating cloud.
It is likely that the B1-b core consists of two velocity components, each of which harbors a very young source.

\end{abstract}


\section{INTRODUCTION}
\indent
The object formed at the beginning of the star-formation process, the first hydrostatic core, was theoretically predicted by \citet{Larson69}. Since the first hydrostatic core is in the transient stage between starless and protostellar cores, it is a key object to understand how stars are formed in molecular clouds.
In the spherical symmetric case, first hydrostatic cores are extremely short-lived, making the expected number of them only few hundredth of Class 0 protostars \citep{Omukai_07, Masun_00}. On the other hand, recent numerical simulations have shown that the rotating first cores can survive longer \citep{Saigo08, Saigo_06}, providing more opportunities to observe them. Lately, some candidates for first cores have been reported \citep[e.g.,][]{Chen_10, Enoch_10, Pineda_11, Chen_12}; they have  extremely low luminosity of $<$ 0.1 L$_\sun$ and are not detectable at wavelengths shorter than $\sim$ 24 $\mu$m. However, the nature of these newly discovered sources have not been well understood yet.

\indent
The chemical properties are also useful tools to characterize the evolutionary stage of prestellar and protostellar cores.
In the cold ($\sim$ 10 K) and dense ($\sim$ 10$^4$ cm$^{-3}$) environment, carbon--bearing species such as CO and CS condense onto dust grains \citep{Caselli02, Taf_02}. In contrast, nitrogen--bearing species remain in the gas phase longer \citep{Bergin_97, Bello_04}. It is known that N$_2$H$^+$ and N$_2$D$^+$ lines are considered to be good tracers of cold dense gas because their abundances increase considerably if the CO is depleted. Observations of prestellar cores show that the spacial distributions of N$_2$H$^+$ and N$_2$D$^+$ well agree with that of dust continuum emission. In addition, in the cold and dense environment, the abundance of deuterium isotopologues is known to be enhanced significantly as compared to the elemental D/H ratio of $\sim$ 1.5 $\times$ 10$^{-5}$ \citep{Oli_03}. The N$_2$D$^+$/N$_2$H$^+$ ratio derived in a sample of prestellar cores ranges from 0.05 to 0.4 \citep{Crapsi05, Gerin01}. There is a tight correlation between N$_2$D$^+$/N$_2$H$^+$  ratio and the CO depletion factor \citep{Crapsi05}. Once a protostar is formed in the core, the protostar warms up its environment and starts evaporating CO into the gas phase. As a result, the N$_2$D$^+$/N$_2$H$^+$ ratio drops as the protostar evolves \citep{Emp09}. 
These results suggest that the N$_2$D$^+$/N$_2$H$^+$  ratio can be used as an indicator of evolution in both prestellar and protostellar cores. Especially protostellar cores with high N$_2$D$^+$/N$_2$H$^+$  ratio are particularly important because they are considered to be in a stage shortly after the formation of the central source.

\indent
In this study, we present the N$_2$H$^+$ $J$=3-2 and N$_2$D$^+$ $J$=3-2 observations with the Submillimeter Array (SMA)\footnote{The Submillimeter Array (SMA) is a joint project between the Smithsonian Astrophysical Observatory and the Academia Sinica Institute of Astronomy and Astrophysics and is funded by the Smithsonian Institute and the Academia Sinica.}
 and the Submillimeter Telescope (SMT) toward two sources embedded in the Barnard 1 (B1) dark cloud \citep[d=250$\pm$50 pc;][]{Cer_03, Hirota_08}, which is part of the Perseus molecular cloud complex. These two sources, labeled as Barnard 1-bN (B1-bN) and Barnard 1-bS (B1-bS) by \citet{Hirano_99}, have no counterparts in the Spitzer images taken by the "From Molecular Cores to Planet Forming Disks" \citep[c2d;][]{Evans03} survey in all four bands of IRAC, and 24 $\mu$m, 70 $\mu$m bands of MIPS \citep{Jo04,Re07}. Both sources reveal the similar SEDs, which show extremely low dust temperatures of $<$ 20 K \citep{Hirano:2012}.  Recently, \citet{Pezzuto:2012} detected the far-infrared emission from B1-bN and B1-bS using the Herschel Space Observatory. On the basis of the SEDs, they proposed that B1-bN and B1-bS are first core candidates.\\

\section{OBSERVATIONS}
\label{sec:obs}

\subsection{Millimeter-interferometer Observations --- SMA}
\indent
SMA observations of the N$_2$H$^+$ $J$=3-2 line were made on 2008 September 3. The array was in the subcompact configuration with seven antennas. The primary-beam of the 6 m diameter antennas at 279 GHz was measured to be $\sim$ 45$\arcsec$. The phase tracking center of the N$_2$H$^+$ $J$=3-2 observations was at $\alpha$(2000) = 3$\arcdeg$33$\arcmin$21$\farcs$4 and $\delta$(2000) = 31$\arcdeg$07$\arcmin$35$\farcs$3. The spectral resolution mode was set to 101.6 kHz (1024 channels per 104 MHz chunk) for the N$_2$H$^+$ $J$=3-2. The corresponding velocity resolution is 0.11 km s$^{-1}$.
A pair of nearby quasars, 3C84 and 3C111, were used for gain calibration. Uranus was used for bandpass and absolute flux calibrations.

\indent
Observations of the N$_2$D$^+$ $J$=3-2 line were carried out on 2007 September 10$\rm^{th}$ and 11$\rm^{th}$. The array was in the subcompact configuration. The primary-beam size of the antenna at 230 GHz was measured to be $\sim$ 54$\arcsec$. The phase tracking center was at the same position as the N$_2$H$^+$ observations. 
The frequency resolution was set to be 101.6 kHz, which corresponds to a velocity resolution of 0.13 km s$^{-1}$.
 The N$_2$D$^+$ $J$=3-2 line was observed simultaneously with the CO $J$=2-1, $\rm^{13}$CO $J$=2-1, C$\rm^{18}$O $J$=2-1, and 230 GHz continuum, which will be presented in the separate paper by \citet{Hirano:2012}.
 
\indent
The visibility data were calibrated and edited using the MIR software package. The calibrated visibility data were Fourier--transformed and CLEANed with MIRIAD \citep{Sault95}, using natural weighting. The synthesized beam sizes were 4$\farcs$53  $\times$ 3$\farcs$21 with a position angle (P.A.) of 44.0$\arcdeg$ for the N$_2$H$^+$ $J$=3-2 and 6$\farcs$08  $\times$ 3$\farcs$17 with a P.A. of 49.2$\arcdeg$ for the N$_2$D$^+$ $J$=3-2.

\indent
The 1.1 mm continuum data were obtained by averaging the line-free channels of the N$_2$H$^+$ $J$=3-2 observations. The 1.1 mm continuum map had a resolution of 4$\farcs$53  $\times$ 3$\farcs$21, and  the rms noise level of $\sim$ 0.35 mJy beam$^{-1}$.

\subsection{Single-dish Observations --- SMT}
\indent
Observations of the N$_2$H$^+$ $J$=3-2 and N$_2$D$^+$ $J$=3-2 lines toward B1-b region were undertaken with the Arizona Radio Observatory sub-millimeter Telescope (SMT) in 2008 November and December. In both lines, we made a 9 $\times$ 9 point map with a grid spacing of 10$\arcsec$. The map covers an area of 1.5$\arcmin$ $\times$ 1.5$\arcmin$ in right ascension and declination centered at $\alpha$(2000) = 3$\arcdeg$33$\arcmin$56$\arcsec$ and $\delta$(2000) = 31$\arcdeg$09$\arcmin$34$\farcs$8. We used the 1.3 mm receiver system equipped with sideband separating mixers. The main-beam efficiency of the telescope was $\sim$ 0.74. The half-power beam widths (HPBW) were 27$\arcsec$ and 33$\arcsec$ in the N$_2$H$^+$ $J$=3-2 and N$_2$D$^+$ $J$=3-2 observations, respectively. For the N$_2$H$^+$ $J$=3-2 observations, we used the chirp transform spectrometer (CTS), which provided a spectral resolution of 0.04 MHz. On the other hand, the N$_2$D$^+$ $J$=3-2 spectra were obtained using the filter bank spectrometers, the spectral resolution of which was 0.25 MHz.

\subsection{Combined Interferometric and Single-Dish Observations}
\indent
We combined the SMT data with the SMA data in order to fill the short-spacing information that was not sampled by the interferometer. We followed the procedure described by \citet{Yen11}, \citet{Taka07} and \citet{Wilner94}. The combination of the data and subsequent imaging were made using the MIRIAD software package.

\indent
The single dish image cube was Fourier transformed into the visibility data. Then the visibility data made from the single-dish image cube and those observed with the SMA were inverse Fourier transformed simultaneously into the image plane with natural weighting. The SMA+SMT combined image cube was made with a velocity interval of 0.3 km s$^{-1}$. The synthesized beam sizes of the combined maps were 4$\farcs$21  $\times$ 3$\farcs$83 with a P.A of 49.81$\arcdeg$ for the N$_2$H$^+$ $J$=3-2 and 6$\farcs$40  $\times$ 4$\farcs$05 with a P.A of 45.75$\arcdeg$ for the N$_2$D$^+$ $J$=3-2. The rms noise level of the combined N$_2$H$^+$ data cube was 0.17 Jy beam$^{-1}$ and that of the N$_2$D$^+$ data cube was 0.13 Jy beam$^{-1}$.

\begin{figure}
\epsscale{.80}
\includegraphics[width=8.5cm]{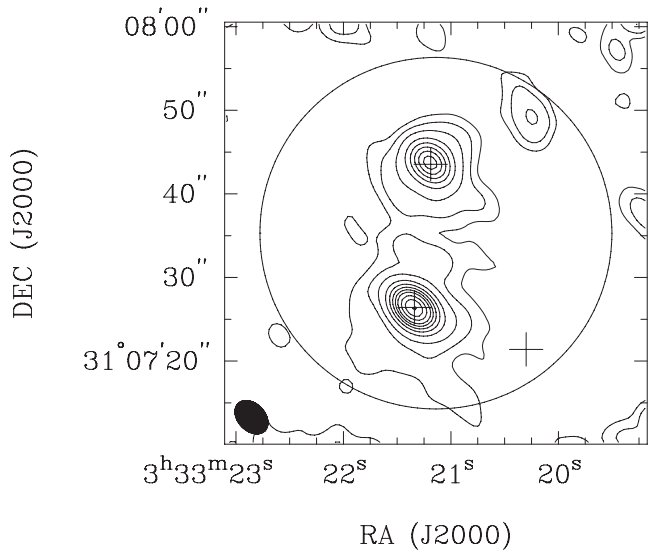}
\caption{The 1.1 mm dust emission observed with the SMA. The contour levels are -3 ,3 ,6 ,12 ,24 ,36 ,48 ,60 ,72 ,90 ,108 ,126 ,144 times the 1$\sigma$ sensitivity of 3.5 mJy beam$^{-1}$. The crosses indicate the peaks of 1.1 mm continuum emission and the peak of the spitzer source. The SMA primary beam size is shown as the large circle.}
\label{fig:dust}
\end{figure}

\section{RESULTS}
\label{sec:Results}
\subsection{Dust Continuum Emission}
\label{sec:Dust}
\indent 
Figure \ref{fig:dust} shows the 1.1 mm continuum image of the B1-b region after the primary beam correction. Two sources, B1-bN and B1-bS, separated by $\sim$ 20$\arcsec$ are clearly seen.
The peak flux densities, total integrated fluxes and peak positions 
of B1-bN and B1-bS were determined by two dimensional gaussian fitting.
The peak flux density at B1-bN is measured to be 0.28$\pm$0.03 Jy beam$^{-1}$ and that at B1-bS is 0.45$\pm$0.04 Jy beam$^{-1}$. The beam deconvolved size of B1-bN is 2$\farcs$2 $\times$ 2$\farcs$0 (550 AU $\times$ 500 AU), and that of B1-bS is 2$\farcs$0 $\times$ 1$\farcs$2 (500 AU $\times$ 300 AU). The total fluxes of B1-bN and B1-bS are 0.37 Jy and 0.52 Jy, respectively.
The positions of these compact sources are $\alpha$(2000) = 3$\rm ^h$33$\rm ^m$21.34$\rm ^s$, $\delta$(2000) = 31$\arcdeg$07$\arcmin$26$\farcs$4 and $\alpha$(2000) = 3$\rm ^h$33$\rm ^m$21.19$\rm ^s$, $\delta$(2000) = 31$\arcdeg$07$\arcmin$43$\farcs$6, respectively. 
The primary beam of the SMA also included the mid infrared source located at $\sim$12$\arcsec$ west of B1-bS.
Although this source was detected in all four bands of IRAC and the 24 and 70 $\mu$m bands in MIPS \citep{Jo06}, there is no counterpart of this source in our 1.1 mm map.
The 3 $\sigma$ upper limit of this source is $\sim$ 10 mJy beam$^{-1}$.
 
\indent
We estimated the total mass of gas and dust in B1-bN and B1-bS by assuming the continuum emission to be optically thin at 1.1 mm, and a gas-to-dust ratio to be 100. The masses are estimated as follows:
\begin{equation}
M=\frac{F_\nu D^2}{\kappa_{\nu}B_{\nu}(T_{\rm dust})}
\end{equation}
where $F_\nu$ is the flux density obtained from continuum emission, D is the source distance, and $\kappa_{\nu}$ is the dust mass opacity derived from $\kappa _\nu$=$\kappa_0$($\nu$/$\nu$$_0$)$^\beta$, where $\kappa _0$ is calculated to be 0.01 cm$^2$ g$^{-1}$ at $\kappa _0$ =231 GHz \citep{Ossen1994}. We adopted the $\beta$ and $T_{\rm dust}$ from the fits of SEDs in \citet{Hirano:2012}. The value of $\beta$ at B1-bN is 1.8$\pm$0.4 and that at B1-bS is 1.3$\pm$0.2. The values of $T_{\rm dust}$ at B1-bN and B1-bS are 15.6$\pm$2.2 K and 18.6$\pm$1.6 K, respectively.
The total masses of B1-bN and B1-bS are calculated to be 0.31$\pm$0.20 M$_\odot$ and 0.34$\pm$0.14 M$_\odot$, respectively. 

\indent
The H$_2$ column densities at B1-bN and B1-bS were also estimated using the same assumptions of $\beta$, $T_{dust}$ and  $\kappa _0$. The resultant $N$(H$_2$) at B1-bN and B1-bS are 2.67$\pm$0.31 $\times$ 10$^{22}$ cm$^{-2}$ and 4.84$\pm$0.72 $\times$ 10$^{22}$ cm$^{-2}$, respectively.

\subsection{Molecular Line Maps}
\label{sec:Line Maps}
\begin{figure*}
\begin{center}
\includegraphics[width=13cm]{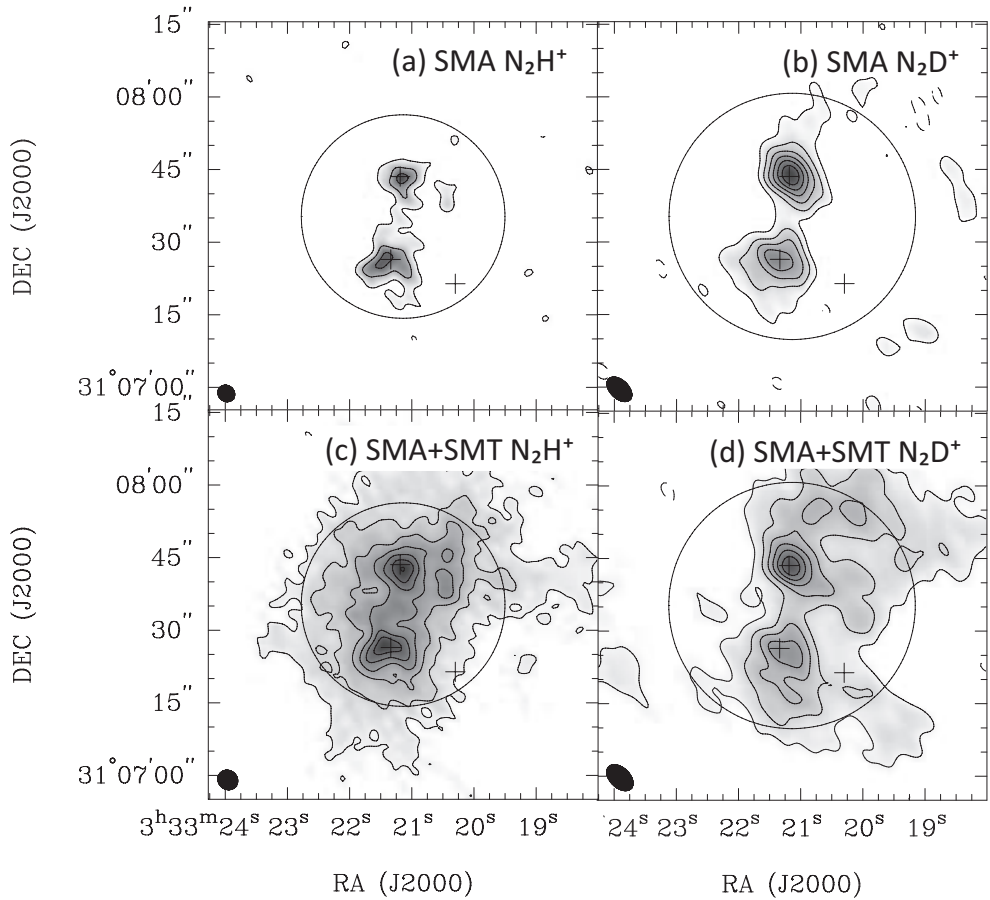}
\epsscale{1.3}
\caption{Integrated line maps of (a) SMA N$_2$H$^+$ $J$=3-2, (b) SMA N$_2$D$^+$ $J$=3-2, (c) combined SMA+SMT N$_2$H$^+$ $J$=3-2, and (d) combined SMA+SMT N$_2$D$^+$ $J$=3-2. The crosses indicate the peaks of 1.1 mm continuum emission and the peak of the spitzer source.
 The black oval on the lower left side and the large circle in each panel show the synthesized beam size and SMA primary beam size, respectively. Contour levels start from 3$\sigma$ by steps of 3$\sigma$ for all figures. The values of 1$\sigma$ are 0.17 Jy beam$^{-1}$, 0.13 Jy beam$^{-1}$, 0.17 Jy beam$^{-1}$ and 0.13 Jy beam$^{-1}$ for (a) to (d), respectively.}
\label {fig:line_mom0}
\end{center}
\end{figure*}

Figure \ref{fig:line_mom0} shows the integrated intensity maps of the N$_2$H$^+$ $J$=3-2 and N$_2$D$^+$ $J$=3-2 lines. The panels at the top of the figure are the maps obtained with only the SMA data, while the ones at the bottom include the contribution of the SMT data. The lines are integrated over the velocity ranges from V$\rm_{LSR}$ = 5.4 -- 8.7 km s$^{-1}$ for the N$_2$H$^+$ $J$=3-2 and from V$\rm_{LSR}$ = 5.7 -- 7.8 km s$^{-1}$ for the N$_2$D$^+$ $J$=3-2. The velocities of N$_2$H$^+$ $J$=3-2 and N$_2$D$^+$ $J$=3-2 are given with respect to the reference rest frequencies of 279.51169 GHz and 231.32990 GHz, respectively. The peaks of N$_2$H$^+$ $J$=3-2 and N$_2$D$^+$ $J$=3-2 emission agree well with those of 1.1 mm continuum emission in both SMA and SMA+SMT maps. In contrast, no significant N$_2$H$^+$ and N$_2$D$^+$ emission was detected at the position of the Spitzer source. 
From the comparisons between the line profiles of the SMA data and that of the combined SMA+SMT data, the missing fluxes of the SMA observations were estimated to be $\sim$ 80 \% and $\sim$ 67 \% for the N$_2$H$^+$ $J$=3-2 and N$_2$D$^+$ $J$=3-2, respectively. The larger missing flux in N$_2$H$^+$ implies that the N$_2$H$^+$ $J$=3-2 emission is more spatially extended than the N$_2$D$^+$ $J$=3-2 emission. 

\begin{figure*}
\begin{center}
\includegraphics[width=13cm]{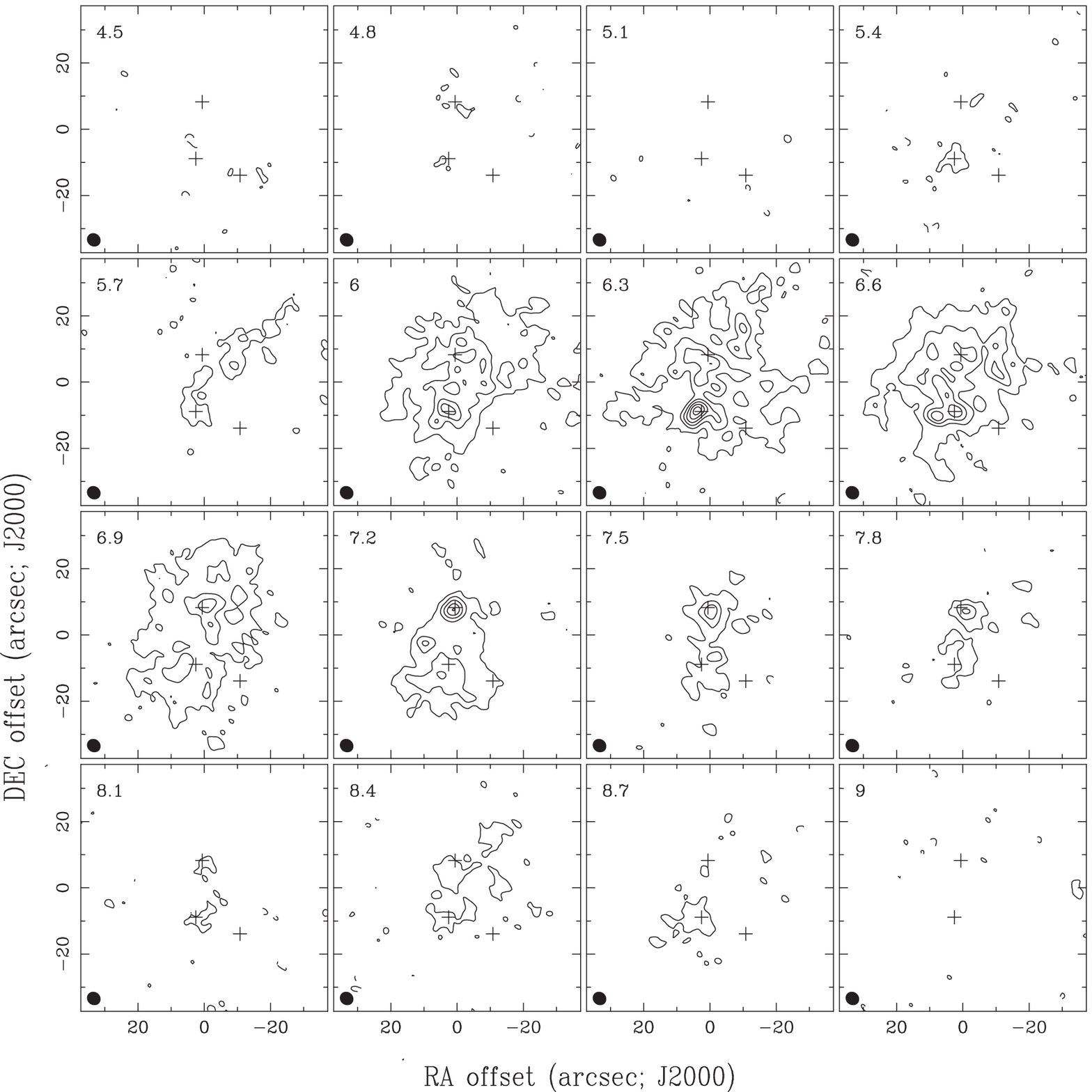}
\epsscale{1}
\caption{Velocity channel maps of the combined SMA+SMT N$_2$H$^+$ $J$=3-2 data in B1-bN and B1-bS. The central velocity at each channel is shown at the top left corner. The small black oval at lower right in each panel indicates the synthesized beam size. The crosses are the peaks of 1.1 mm continuum emission and the peak of the spitzer source. Contour levels start from 3$\sigma$ by steps of 3$\sigma$ for all figures, where 1$\sigma$ is 0.17 Jy beam$^{-1}$.}
\label {fig:n2h_cha}
\end{center}
\end{figure*}

\indent
Figure \ref{fig:n2h_cha} shows the velocity channel maps of the combined SMA+SMT N$_2$H$^+$ $J$=3-2 data. In the low velocity range (V$\rm_{LSR}$ = 6.0 - 6.6 km s$^{-1}$), the N$_2$H$^+$ line emission is detected around both B1-bN and B1-bS regions. In this velocity range, the N$_2$H$^+$ emission reveals a centrally peaked distribution at B1-bS (especially at V$\rm_{LSR}$ $\sim$ 6.3 km s$^{-1}$), while it is more spatially extended around B1-bN. In the high velocity range (V$\rm_{LSR}$ = 6.9 - 7.8 km s$^{-1}$), the emission is centrally peaked at B1-bN, while it is faint and spatially extended at B1-bS.

\begin{figure*}
\begin{center}
\includegraphics[width=13cm]{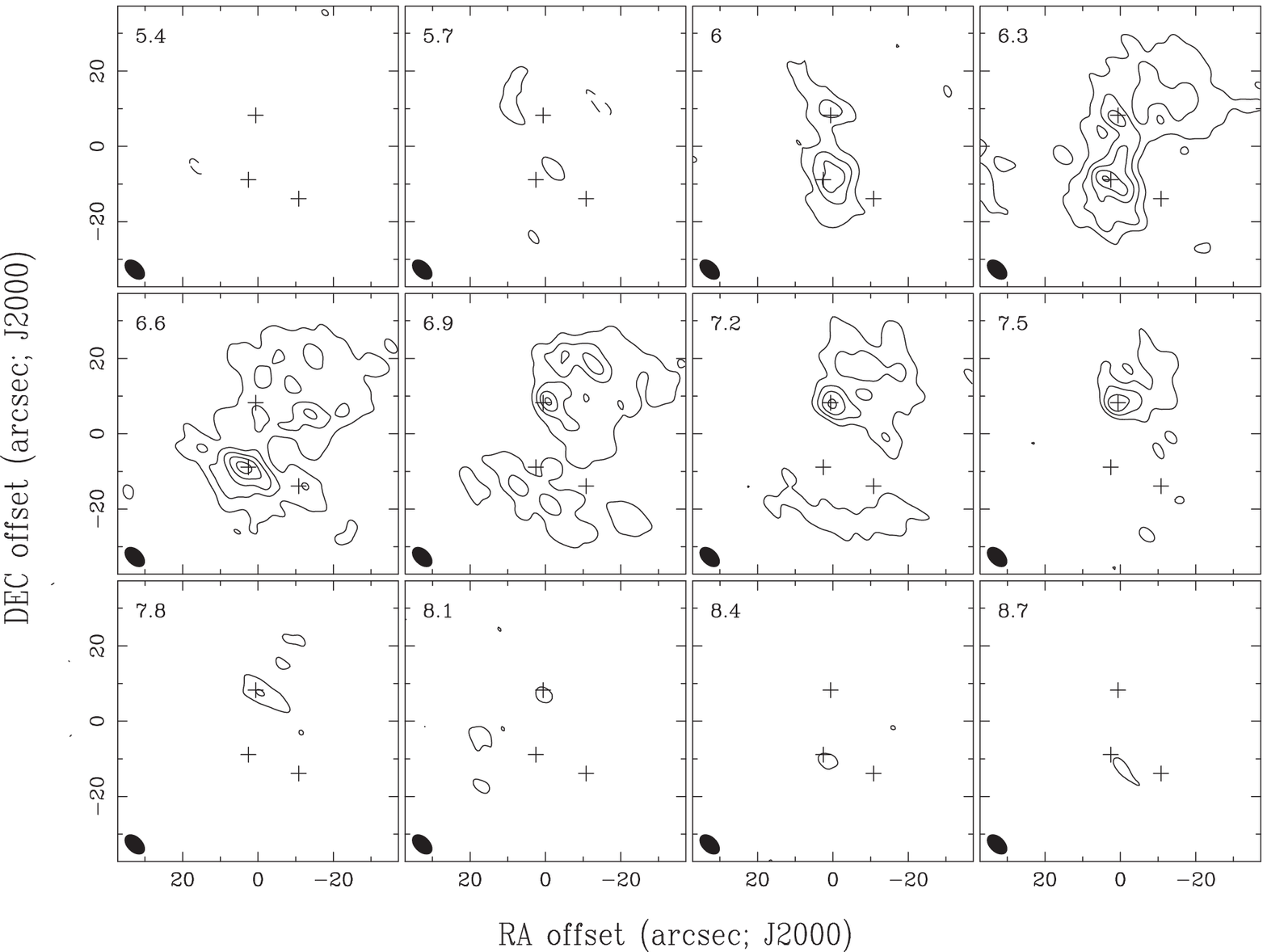}
\epsscale{1}
\caption{Velocity channel maps of the combined SMA+SMT N$_2$D$^+$ $J$=3-2 data in B1-bN and B1-bS. The central velocity at each channel is shown at the top left corner. The small black oval at lower right in each panel indicates the synthesized beam size. The crosses are the peaks of 1.1 mm continuum emission and the peak of the spitzer source. Contour levels start from 3$\sigma$ by steps of 3$\sigma$ for all figures, where 1$\sigma$ is 0.13 Jy beam$^{-1}$.}
\label {fig:n2d_cha}
\end{center}
\end{figure*}

\indent
The velocity channel maps of the combined SMA+SMT N$_2$D$^+$ data (Fig. \ref{fig:n2d_cha}) show the same trend as the N$_2$H$^+$ maps; the compact components at the positions of B1-bN and B1-bS appear in the velocity ranges of 6.9 - 7.8 km s$^{-1}$ and 6.0 - 6.6 km s$^{-1}$, respectively. 
As in the case of the N$_2$H$^+$, the spatially extended N$_2$D$^+$ emission around B1-bN is also seen in the low velocity range.

\begin{figure*}
\begin{center}
\includegraphics[width=11cm]{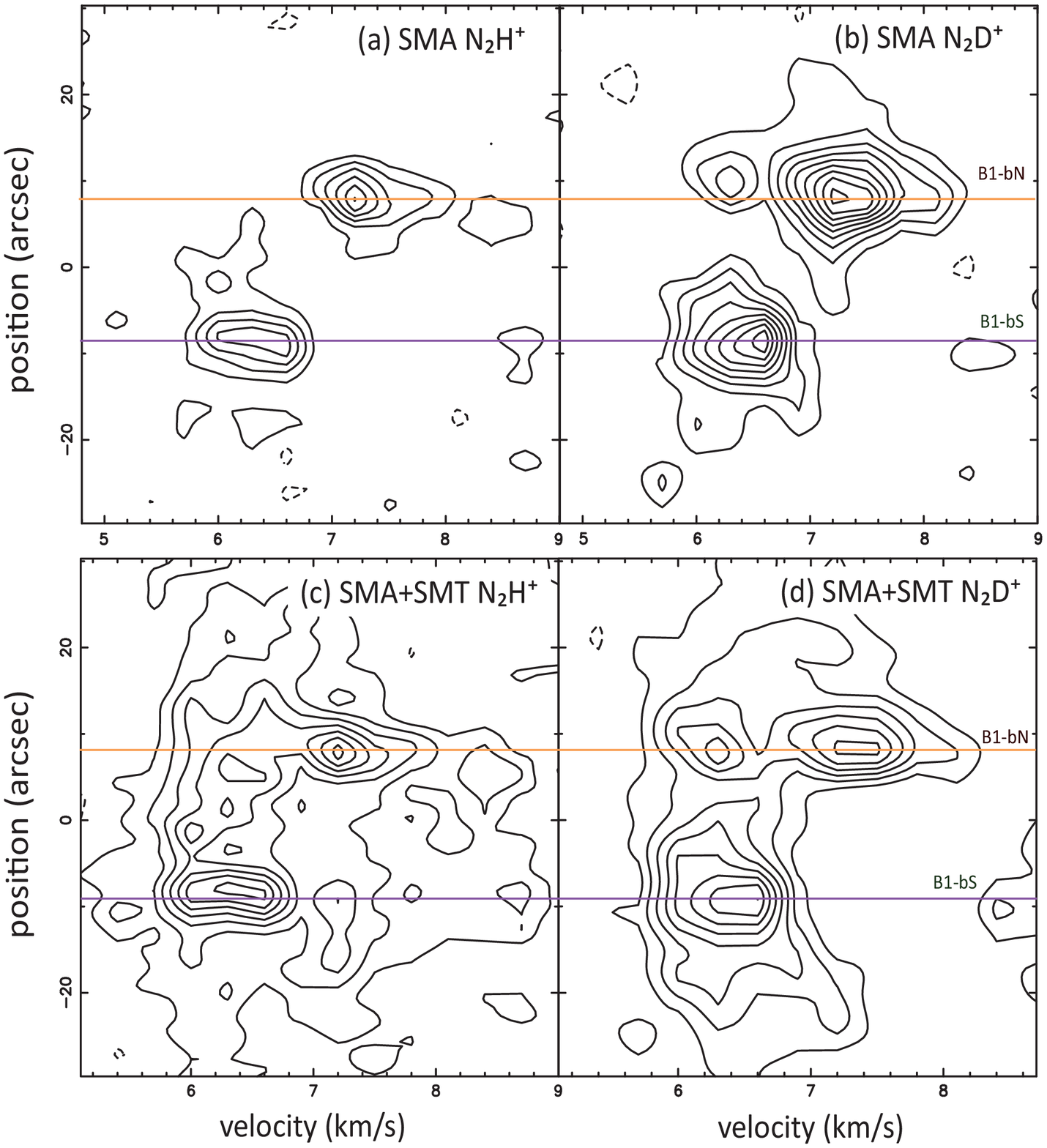}
\epsscale{1}
\caption{P--V diagrams of (a) SMA N$_2$H$^+$ $J$=3-2, (b) SMA N$_2$D$^+$ $J$=3-2, (c) Combined SMA+SMT N$_2$H$^+$ $J$=3-2 and (d) N$_2$D$^+$ $J$=3-2 emission along the N--S cut through B1--bN and B1--bS. The horizontal lines indicate the positions of B1--bN (orange) and B1--bS (purple). Contour levels are from 2$\sigma$ in steps of 2$\sigma$, where 1$\sigma$ are 0.17 and 0.13 Jy beam$^{-1}$ in the N$_2$H$^+$ and N$_2$D$^+$ data, respectively.}
\label{fig_PV}
\end{center}
\end{figure*}

\subsection{Kinematic Structure}
\label{sec:kine}
\indent
In order to examine the velocity structure of B1-b, we made the position-velocity (P--V) diagrams of the N$_2$H$^+$ and N$_2$D$^+$ along the North-South cut through B1-bN and B1-bS (Fig.  \ref{fig_PV}). The upper panels of this figure are the diagrams obtained with the SMA data alone, while the ones in the lower panels include the contribution of the SMT data. The upper panels clearly show two velocity components; one is the 7.2 km s$^{-1}$ component at B1-bN and the other is the 6.3 km s$^{-1}$ component at B1-bS. 
At the position of B1-bN, the N$_2$D$^+$ emission has a secondary component at V$\rm_{LSR}$ $\sim$ 6.3 km s$^{-1}$. 
This component is connected to the 6.3 km s$^{-1}$ component at B1-bS in the SMA+SMT combined map shown in the lower panels.
The SMA+SMT combined map of N$_2$H$^+$ shows the faint 7.2 km s$^{-1}$ component at the position of B1-bS, however, this component is not seen in the N$_2$D$^+$ map.

\section{ANALYSIS}
\label{ref:ana}
\subsection{N$_2$H$^+$  and N$_2$D$^+$ Hyperfine Fitting}
\label{sec:fit}
\indent
Since N$_2$H$^+$ $J$=3-2 and N$_2$D$^+$ $J$=3-2 rotational transitions contain numerous hyperfine components, line parameters such as central velocity (V$_c$), line width ($\bigtriangleup v$), optical depth ($\tau$), excitation temperature (T$\rm_{ex}$) need to be determined by hyperfine fitting. We assume all the hyperfine components to have the same excitation temperature and line width, with the relative velocity of each component fixed to the laboratory value.
Since B1-bN has two velocity components (Fig.  \ref{fig_PV}), it was difficult to determine all the parameters from fitting.
Therefore, we fixed the excitation temperatures to be 7.6 K for N$_2$H$^+$ and 6.7 K for N$_2$D$^+$, which were determined by \citet{Emp09}. Since these excitation temperatures were derived from the N$_2$H$^+$ $J$=1-0 and N$_2$D$^+$ $J$=1-0 single-dish observations with HPBW of $\sim$ 27$\arcsec$ and 32$\arcsec$, respectively, the values are considered to be the average values of the B1-b region. We adopted the N$_2$H$^+$ $J$=3-2 hyperfine components listed in \citet{Caselli02} and those of the N$_2$D$^+$ $J$=3-2 listed in \citet{Gerin01}.
We used the combined SMA+SMT data to perform the fitting. The N$_2$H$^+$ data were convolved in order to match the beam size of the N$_2$D$^+$ data (6$\farcs$40  $\times$ 4$\farcs$05 with a P.A of 45.75\arcdeg). The primary beam of the SMA antennas were also corrected.

\begin{figure*}
\begin{center}
\epsscale{1}
\includegraphics[width=13cm]{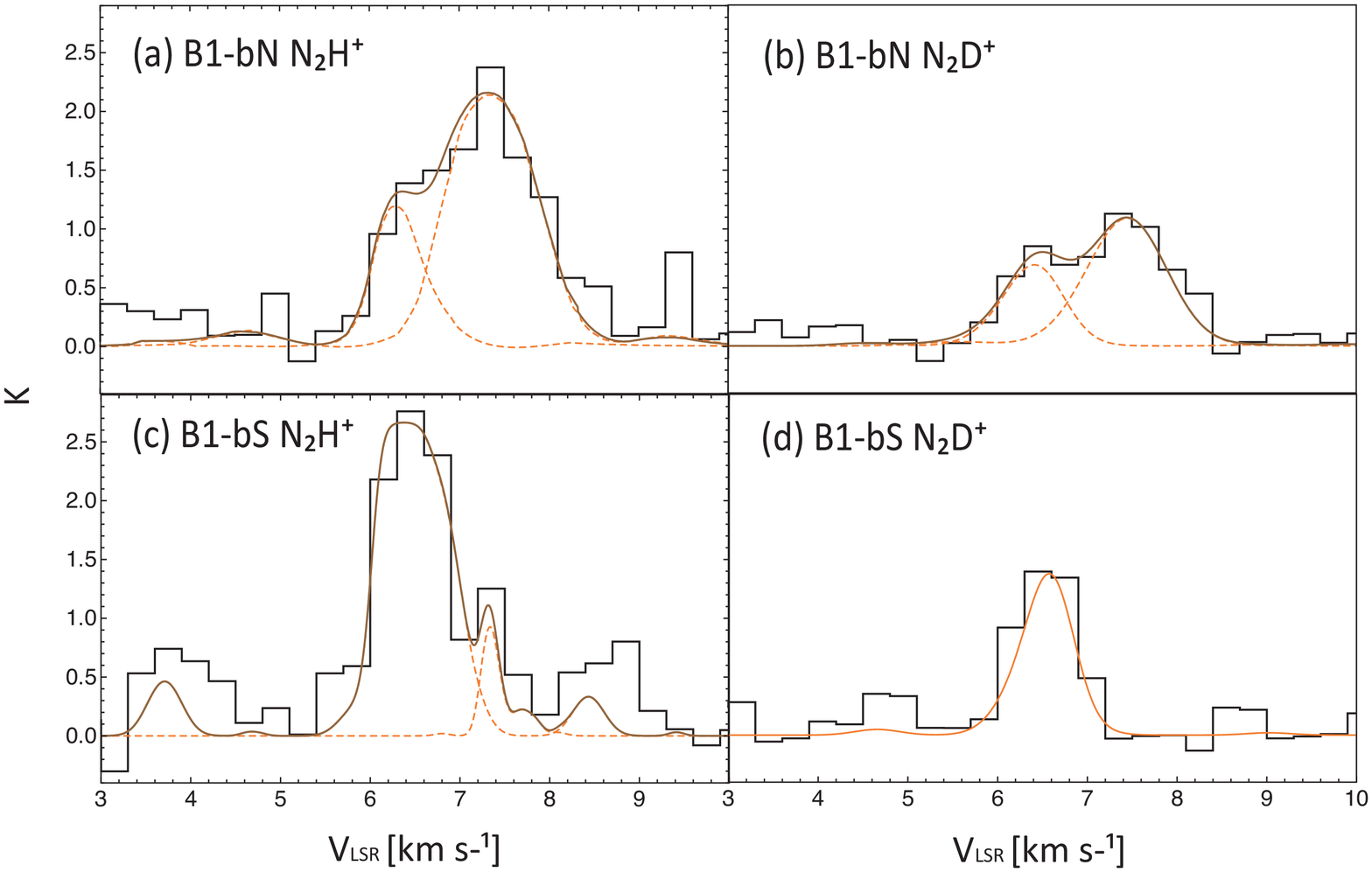}
\caption{Line profiles of (a) N$_2$H$^+$ $J$=3-2 and (b) N$_2$D$^+$ $J$=3-2 at the position of B1-bN, (c) N$_2$H$^+$ $J$=3-2 and (d) N$_2$D$^+$ $J$=3-2 at that of B1-bS.   The dashed curves (orange) indicate individual velocity components, while the solid curves (brown) show the overall fitting results. }
\label{fig:spectra}
\end{center}
\end{figure*}

\indent
Figure \ref{fig:spectra} shows the combined SMA+SMT N$_2$H$^+$ $J$=3-2 and N$_2$D$^+$ $J$=3-2 line profiles and the results of hyperfine fittings at the positions of B1-bN and B1-bS. 
The N$_2$D$^+$ line profiles were fitted well using the optically thin assumption, while the N$_2$H$^+$ lines were obviously optically thick.
The spectra at B1-bN has two velocity components; one is at V $\sim$ 7.2 km s$\rm^{-1}$, which is spatially compact, and the other is the spatially extended component at V $\sim$ 6.2 km s$\rm^{-1}$. The line profiles at B1-bS are mostly dominated by the V$\rm_{LSR}$ = 6.3 km s$\rm^{-1}$ component.  
The determined parameters are listed in Table \ref{tab:parah} and Table \ref{tab:parad}. In these tables, the main components refer to the dominate components at B1-bN and B1-bS, which correspond to the velocity components centered at $\sim$ 7.2 km s$^{-1}$ and 6.3 km s$^{-1}$, respectively.
Since the line profile of N$_2$H$^+$ $J$=3-2 at the position of B1-bS shows the secondary component at $\sim$ 7.2 km s$^{-1}$, we included this component to the fitting.

\indent
It should be noted that the N$_2$H$^+$ line becomes optically thin if the higher excitation temperature is assumed; the optical depth of the primary component of B1-bN becomes less than 1.0 for the T$\rm_{ex}$ $>$ 8.8 K, and that of B1-bS becomes $<$ 1 for the T$\rm_{ex}$ $>$ 10.5 K. On the other hand, the optical depth of the N$_2$H$^+$ line becomes the largest ($\sim$ 10 for B1-bN and $\sim$ 18 for B1-bS) if we adopt the lower limit of the excitation temperature derived from the fitting ($\sim$ 6.4 K for B1-bN and $\sim$ 7.2 K for B1-bS).

\subsection{N$_2$H$^+$  and N$_2$D$^+$ Column Density}
\indent
We estimated the column densities of N$_2$H$^+$ and N$_2$D$^+$ using the method described by \citet{Caselli02}.
The results including column densities of N$_2$H$^+$ and N$_2$D$^+$, and the $N$(N$_2$D$^+$)/$N$(N$_2$H$^+$) ratios at B1-bN and B1-bS are summarized in Table \ref{tab:column}.
The column densities of N$_2$H$^+$ at the positions of B1-bN and B1-bS were calculated to be $\sim$ 1.1 $\times$ 10$\rm^{13}$ cm$\rm^{-2}$ and 1.8 $\times$ 10$\rm^{13}$ cm$\rm^{-2}$, respectively. While column densities of N$_2$D$^+$ at B1-bN and B1-bS are $\sim$ 2.9 $\times$ 10$\rm^{12}$ cm$\rm^{-2}$ and 2.4 $\times$ 10$\rm^{12}$ cm$\rm^{-2}$, respectively. 
The corresponding N$_2$D$^+$/N$_2$H$^+$ ratios was $\sim$ 0.27 at B1-bN and $\sim$ 0.13 at B1-bS. These values are consistent with the N$_2$D$^+$/N$_2$H$^+$ ratio of $\sim$ 0.18 in the entire B1-b region obtained by \citet{Emp09}. 
The N$_2$H$^+$ column density becomes lower if we adopt higher excitation temperatures (a factor of 6 - 9 if T$\rm_{ex}$ = 20 K). However, the N$_2$D$^+$/N$_2$H$^+$ ratio is almost constant at $\sim$ 0.3 - 0.4, because both N$_2$H$^+$ and N$_2$D$^+$ lines are optically thin at T$\rm_{ex}$$\ga$ 10 K. On the other hand, if we adopt the lower limit of the excitation temperature, the column density of N$_2$H$^+$ increases by a factor of $\sim$ 3 - 4. Since the N$_2$D$^+$ line is assumed to be optically thin, the N$_2$D$^+$/N$_2$H$^+$ ratio decreases to $\sim$ 0.06.

\indent
The fractional abundances of N$_2$H$^+$ at B1-bN and B1-bS were calculated to be  3.97$\pm$0.62$\times$10$^{-10}$ and  3.71$\pm$0.63$\times$10$^{-10}$, respectively, using the N$_2$H$^+$ column densities and H$\rm_2$ column densities derived from the continuum data.
\citet{Jo04} derived the abundance of N$_2$H$^+$ in prestellar cores and Class 0 sources to be $\sim$ 10$^{-8}$ -- 10$^{-10}$. The N$_2$H$^+$ abundance in these two sources is comparable to those of the samples studied by \citet{Jo04}.

\begin{deluxetable*}{lccccccc}
\tablecolumns{8}
\tablewidth{0pc}
\tablecaption{Results of N$_2$H$^+$ $J$=3-2 Hyperfine Fitting\label{tab:parah}}
\tablehead{
\colhead{}    &  \multicolumn{3}{c}{Main Component} &   \colhead{}   &
\multicolumn{3}{c}{Second Component} \\
\cline{2-4} \cline{6-8} 
\colhead{Core} & \colhead{$\tau _{tot}$}   & \colhead{V$_c$ (km s$\rm^{-1}$)}    & \colhead{$\bigtriangleup v$ (km s$\rm^{-1}$)} &
\colhead{}    & \colhead{$\tau _{tot}$}   & \colhead{V$_c$ (km s$\rm^{-1}$)}    & \colhead{$\bigtriangleup v$ (km s$\rm^{-1}$)}}
\startdata
B1-bN & 2.00$\pm$0.03 & 7.19$\pm$0.02 & 0.83$\pm$0.05 &
& 0.82$\pm$0.02 & 6.18$\pm$0.02 & 0.50$\pm$0.05 \\
B1-bS & 8.00$\pm$0.06 & 6.27$\pm$0.02 & 0.35$\pm$0.03 &&\tablenotemark{a} 0.10$\pm$0.08 
 & 7.25$\pm$0.01 & 0.21$\pm$0.02 
\enddata
\tablenotetext{a}{Though the second component is not clear at B1-bS, two components were used to fit the spectrum of N$_2$H$^+$ $J$=3-2 at B1-bS because it seems to be a more spatially-extended component centered at $\sim$ 6.2 km s$^{-1}$.}
\end{deluxetable*}

\begin{deluxetable*}{lccccccc}
\tablecolumns{8}
\tablewidth{0pc}
\tablecaption{Results of N$_2$D$^+$ $J$=3-2 Hyperfine Fitting\label{tab:parad}}
\tablehead{
\colhead{}    &  \multicolumn{3}{c}{Main Component} &   \colhead{}   &
\multicolumn{3}{c}{Second Component} \\
\cline{2-4} \cline{6-8} 
\colhead{Core} & \colhead{I\tablenotemark{a} (K)}  & \colhead{V$_c$ (km s$\rm^{-1}$)}    & \colhead{$\bigtriangleup v$ (km s$\rm^{-1}$)} &
\colhead{}    & \colhead{I (K)}   & \colhead{V$_c$ (km s$\rm^{-1}$)}    & \colhead{$\bigtriangleup v$ (km s$\rm^{-1}$)} }
\startdata
B1-bN & 1.30$\pm$0.06 & 7.33$\pm$0.02 & 0.94$\pm$0.05 &
& 0.87$\pm$0.08 & 6.28$\pm$0.03 & 0.69$\pm$0.04 \\
B1-bS & 1.80$\pm$0.03 & 6.43$\pm$0.03 & 0.58$\pm$0.05 &&  \nodata  
 &  \nodata  &  \nodata 
\enddata
\tablenotetext{a}{The brightness temperature of each hyperfine component is I $\times$ (relative intensity of each component).}
\end{deluxetable*}

\section{DISCUSSIONS}
\label{sec:dis}
\subsection{Deuterium Fractionation}
\label{sec:deu}
\indent
The N$_2$D$^+$/N$_2$H$^+$ ratio of 0.13 -- 0.27 derived in the previous section is 4 orders of magnitude larger than the D/H ratio of 10$^{-5}$ in the interstellar space \citep{Oli_03}. It is known that the deuterium fractionation in N$_2$H$^+$ is related to the core temperature and CO depletion \citep[e.g.,][] {Fran07}. In the proton-deuteron exchange reaction, H$_3^+$+HD$\rightleftharpoons$H$_2$D$^+$+H$_2$ \citep{Millar89},
H$_2$D$^+$/H$_3^+$ tends to be increased at lower temperature. In addition, at densities of $\ga$ 10$^5$ cm$^{-3}$, the depletion of carbon-bearing molecules onto dust grains further increases the H$_2$D$^+$/H$_3^+$ ratio \citep{RM20}. 
Since H$_2$D$^+$ is the main ingredient of deuteriated molecules, the enhancement of H$_2$D$^+$ increases the abundance of deuteriated molecules.
Therefore, the N$_2$D$^+$/N$_2$H$^+$ ratio becomes an indicator of the physical conditions and molecular depletion of dense and cold gas, and is useful to probe the chemical and dynamical evolution of cores, including starless cores and class 0 protostars. 

\begin{deluxetable*}{lccccccc}
\tablecolumns{7}
\tablewidth{0pc}
\tablecaption{N$_2$H$^+$, N$_2$D$^+$ and H$_2$ column densities\label{tab:column}}
\tablehead{
\colhead{Core} &\colhead{}  & \colhead{$N$(N$_2$H$^+$) (cm$^{-2}$)}   & \colhead{$N$(N$_2$D$^+$) (cm$^{-2}$)}    & \colhead{$N$(N$_2$D$^+$)/$N$(N$_2$H$^+$)}  & \colhead{$N$(H$_2$) (cm$^{-2}$)}   & \colhead{$N$(N$_2$H$^+$)/$N$(H$_2$)} }
\startdata
B1-bN && 1.06$\pm$0.11$\times$10$^{13}$ & 2.85$\pm$0.09$\times$10$^{12}$ & 0.27$\pm$0.11 & 2.67$\pm$0.31$\times$10$^{22}$ & 3.97$\pm$0.62$\times$10$^{-10}$ \\
B1-bS && 1.80$\pm$0.15$\times$10$^{13}$ & 2.40$\pm$0.09$\times$10$^{12}$ & 0.13$\pm$0.09 & 4.84$\pm$0.72$\times$10$^{22}$ & 3.71$\pm$0.63$\times$10$^{-10}$ 
\enddata
\end{deluxetable*}

\indent
\citet{Crapsi05} observed 31 low-mass starless cores and derived the N$_2$D$^+$/N$_2$H$^+$ ratio to be 0.03 -- 0.44.
They found a strong correlation between the N$_2$D$^+$/N$_2$H$^+$ ratio and CO depletion, suggesting that the N$_2$D$^+$/N$_2$H$^+$ ratio increases with the density of the core.
The higher N$_2$D$^+$/N$_2$H$^+$ values were seen in the cores of later evolutionary stage, i.e. those with centrally concentrated density distribution and high H$_2$ column densities.
In the case of the Class 0 protostellar cores studied by \citet{Emp09}, the N$_2$D$^+$/N$_2$H$^+$ ratio also ranges from 0.03 to 0.3. 
Their study showed that the N$_2$D$^+$/N$_2$H$^+$ ratio in the prestellar cores significantly decreases as a function of dust temperature, suggesting that the ratio decreases as the central star evolves and heats the surrounding gas. 
These results suggest that the largest N$_2$D$^+$/N$_2$H$^+$ ratio is expected to be in prestellar cores that are close to form protostars or protostellar cores in the earliest evolutionary stage.
The N$_2$D$^+$/N$_2$H$^+$ ratios derived at B1-bN and B1-bS ($\sim$ 0.2) are comparable to those of the evolved prestellar cores such as L1544 and the very young protostars such as L1448C, IRAS 03282 and HH211.
It is notable that the column densities of N$_2$D$^+$ estimated here are the lower limits, because we assumed the N$_2$D$^+$ line to be optically thin.
Therefore, the N$_2$D$^+$/N$_2$H$^+$ ratios derived here are also the lower limits.

\indent
The high D/H ratios in the B1-b region are also shown in other observations \citep{Roueff:2002, Roueff_2005, Marcelino_2005}. The high D/H ratios are consistent with the low temperatures derived from the SEDs of the continuum emission \citep{Pezzuto:2012, Hirano:2012}.
Since both B1-bN and B1-bS contain compact continuum sources that are clearly detected by the interferometer, they are unlikely to be the prestellar cores. On the basis of the properties of B1-bN and B1-bS derived from the continuum SEDs such as low temperatures (T$\rm_{dust} <$ 20 K), low intrinsic luminosities (L$\rm_{int} \sim$ 0.03 L$_\odot$ for B1-bS and $\la$ 0.004 L$_\odot$ for B1-bN), and low velocity CO outflows \citep{Hirano:2012}, these two sources are considered to be the candidates for the first hydrostatic cores.

\subsection{Kinematics}
\indent
As shown in Section \ref{sec:kine}, B1-bN appears at V$\rm_{LSR}$ = 7.2 km s$^{-1}$ and B1-bS reveals at V$\rm_{LSR}$ = 6.3 km s$^{-1}$. In other words, two sources with a projected separation of $\sim$ 5000 AU have a velocity difference of $\sim$ 0.9 km s$^{-1}$. If this velocity difference attributes to the orbital motion of gravitationally-bounded objects, the mass of each source is estimated to be $\sim$ 0.55 M$_\odot$. This mass agrees with the mass derived from the continuum data within a factor of two. 
If the two sources originate from a single rotating cloud, the P--V diagram along the line connecting two sources (i.e. in the N--S direction) is expected to show a velocity gradient.
However, the velocity pattern seen in the P--V maps including the short spacing data (Fig. \ref{fig_PV}(c) and Fig. \ref{fig_PV}(d)) does not show such a velocity gradient. Instead, these P--V maps reveal a C-shaped pattern with two velocity components along the line of sight of B1-bN. 
The spatially extended emission connects B1-bS and the secondary component of B1-bN, which has the same velocity as B1-bS.
In addition, the N$_2$H$^+$ map shows an additional 7.2 km s$^{-1}$ component along the line of sight of B1-bS. Therefore, it is unlikely that the two sources were formed in a single rotating cloud through the fragmentation.
The P--V maps including the short spacing data suggest that there are two velocity components along the line of sight of the B1-b core, and that B1-bN belongs to the 7.2 km s$^{-1}$ cloud and B1-bS to the 6.3 km s$^{-1}$ cloud. 

\begin{figure}
\begin{center}
\epsscale{0.8}
\includegraphics[width=7cm]{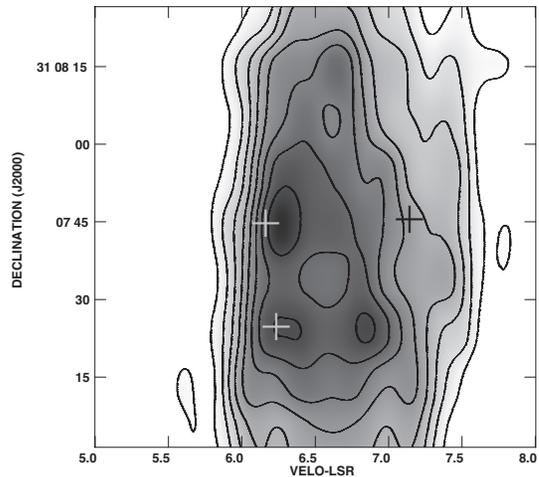}
\caption{P--V diagram of H$^{13}$CO$^+$ $J$=1-0 emission along the N--S cut through B1-bN and B1-bS observed with the NRO 45 m telescope. The crosses represent the peaks of the 6.3 km s$^{-1}$ component in B1-bS and the 6.3 and 7.2 km s$^{-1}$ components in B1-bN. The contour levels are 3, 5, 7, 9, 11, 13, 15, 17, 19, 21 times the 1$\sigma$ sensitivity of 0.07 K.}
\label{fig:pv_H13}
\end{center}
\end{figure}

\subsection{Comparison with H$^{13}$CO$^+$ $J$=1-0 Emission}
\indent
Fig. \ref{fig:pv_H13} shows the P--V diagram of H$^{13}$CO$^+$ $J$=1-0 emission along the N--S cut through B1-bN and B1-bS observed within the NRO 45 m telescope \citep{Hirano:2012}.
Two velocity components along the line of sight of B1-b are also seen in the H$^{13}$CO$^+$ map.
It is notable that the relative intensities of these two velocity components in the H$^{13}$CO$^+$ map are considerably different from those of the N$_2$H$^+$ and N$_2$D$^+$ maps.
As shown in the previous section, the dense gas around B1-bN traced by the N$_2$H$^+$ and N$_2$D$^+$ lines has a velocity of 7.2 km s$^{-1}$. In addition to this 7.2 km s$^{-1}$ component, there is a secondary component at 6.3 km s$^{-1}$. On the other hand, the H$^{13}$CO$^+$ line peaks at 6.3 km s$^{-1}$ and shows no significant enhancement at V$\rm_{LSR}\sim$ 7.2 km s$^{-1}$. 
At B1-bS, both N$_2$H$^+$ and N$_2$D$^+$ maps reveal the 6.3 km s$^{-1}$ component. Although the N$_2$H$^+$ line shows a weak secondary peak at $\sim$ 7 km s$^{-1}$, there is no counterpart of that velocity component in the N$_2$D$^+$ line. In the case of H$^{13}$CO$^+$ map, however, it shows double peaks at 6.3 km s$^{-1}$ and 6.9 km s$^{-1}$. The comparison shows that N$_2$H$^+$ and N$_2$D$^+$ emission is partially anti-correlated with the H$^{13}$CO$^+$ emission.

\indent
These results can be well explained by the different chemical properties between carbon-bearing molecules and nitrogen-bearing molecules. 
Since H$^{13}$CO$^+$ (HCO$^+$) is considered to be formed from $^{13}$CO (CO) through the ion-neutron reaction in the gas phase, the abundance of H$^{13}$CO$^+$ is expected to decrease if a significant fraction of CO molecules freeze onto the dust grains.
In such a cold and dense region with CO depletion, the N$_2$H$^+$ and N$_2$D$^+$ abundance tend to be enhanced, since CO is the main reactant to destroy the N$_2$H$^+$ and N$_2$D$^+$.
Therefore, the chemical properties of the dense gas surrounding B1-bN with a velocity of 7.2 km s$^{-1}$ can be explained by the scenario of CO depletion.
In the case of B1-bS, the H$^{13}$CO$^+$ coexists with N$_2$H$^+$ and N$_2$D$^+$. This is probably because the gas in B1-bS is heated by the newly formed star.
As a consequence, the CO in this region has been partially come back to the gas phase.
The presence of CO outflows around this source also support this idea \citep{Hirano:2012}.
Since H$^{13}$CO$^+$ is made from the reaction N$_2$H$^+$ + CO $\rightleftharpoons$ H$^{13}$CO$^+$ +N$_2$, the existence of N$_2$H$^+$ and CO makes H$^{13}$CO$^+$.
A comparison between B1-bN and B1-bS suggests that B1-bN is chemically younger than B1-bS.
The SEDs in \citet{Hirano:2012} and \citet{Pezzuto:2012} also show that B1-bN has lower dust temperature. 
In addition, B1-bN was not detected at 70 $\mu$m by Herschel (Pezzuto et al. 2012), while B1-bS
was detected in the same wavelength. These results also support the idea that two sources are in slightly different evolutionary stages.

\indent
The 6.3 km s$^{-1}$ component along the line of sight of B1-bN is bright in H$^{13}$CO$^+$ while faint in N$_2$H$^+$ and N$_2$D$^+$, suggesting that  the CO depletion in this velocity component is not significant. On the other hand, the 6.9 km s$^{-1}$ component at  B1-bS is also bright in the H$^{13}$CO$^+$, but unclear in the N$_2$H$^+$ and N$_2$D$^+$. Although the LSR velocity of this component (6.9 km s$^{-1}$) is slightly different from 7.2 km s$^{-1}$ in the N$_2$H$^+$, the 6.9 km s$^{-1}$ component seen in the H$^{13}$CO$^+$ is probably related to the 7.2 km s$^{-1}$ component observed in the N$_2$H$^+$. The chemical properties suggest that the density of the 6.3 km s$^{-1}$ component at B1-bN is higher than that of the  7.2 km s$^{-1}$ component at B1-bS.

\section{CONCLUSIONS}
\indent 
We have observed the two sources in the B1-b core, B1-bN and B1-bS, in 1.1 mm continuum emission with the SMA, and N$_2$H$^+$ $J$=3-2 and N$_2$D$^+$ $J$=3-2 line emission with the SMA and the SMT.
The main conclusions are the followings:
\begin{enumerate}[1. ]
\item Two compact sources, B1-bN and B1-bS, with a projected separation of $\sim$ 20$\arcsec$, are shown in the 1.1 mm continuum map observed with the SMA. Their masses were estimated to be $\sim$ 0.3 M$_{\odot}$. 
\item
The N$_2$H$^+$ and N$_2$D$^+$ lines show significant enhancement at the positions of B1-bN and B1-bS.

\item
The N$_2$H$^+$ abundance was estimated to be $\sim$ 4 $\times$ 10$^{-10}$ at both B1-bN and B1-bS. The derived N$_2$H$^+$ abundance is comparable to those of prestellar cores and Class 0 protostars.

\item
The N$_2$D$^+$/N$_2$H$^+$ ratio was estimated to be $\sim$ 0.2 at both B1-bN and B1-bS. 
This value is 4 orders of magnitude higher than that in the interstellar medium. The N$_2$D$^+$/N$_2$H$^+$ ratio is comparable to those in the late prestellar phase or in the earliest protostellar phase. Since these two sources harbor compact components, they are more likely to be in the very early stage of protostellar formation. 

\item The radial velocities of B1-bN and B1-bS are V$\rm _{LSR}$=7.2 km s $^{-1}$ and V$\rm _{LSR}$=6.3 km s $^{-1}$, respectively. The C-shaped velocity pattern seen in the P--V diagrams of N$_2$H$^+$ and N$_2$D$^+$ does not support the idea that two sources are formed by means of fragmentation of a single rotating cloud. It is likely that there are two velocity components at V$\rm _{LSR}$=6.3 km s $^{-1}$ and 7.2 km s $^{-1}$ along the line of sight of B1-b, each of which harbors a young protostar.

\item At the position of B1-bN, the 7.2 km s$^{-1}$ component, which is bright in N$_2$H$^+$ and N$_2$D$^+$, is not clearly seen in the H$^{13}$CO$^+$ $J$=1-0. This suggests that the carbon bearing molecules in the dense gas surrounding B1-bN are heavily depleted onto the dust.

\item
On the other hand, the 6.3 km s$^{-1}$ component at B1-bS is observed in N$_2$H$^+$, N$_2$D$^+$ and H$^{13}$CO$^+$ lines. The difference in chemical property suggests that B1-bN is in the earlier evolutionary stage as compared to B1-bS.

\end{enumerate}

\indent
We wish to thank all the SMA staff in Hawaii, Cambridge, and Taipei for their kind help during these observations. N. Hirano is supported by NSC grant 99-2112-M-001-009-MY3.

\end{document}